\documentclass[prb,twocolumn,amsmath,showpacs]{revtex4}

\usepackage{graphicx}

\begin{document}
\input epsf


\title{Electron-phonon interaction in transition metal diborides
$T$B$_{2}$ ($T$=Zr, Nb, Ta) studied by point-contact spectroscopy}

\author{Yu. G. Naidyuk$^1$\email{naidyuk@ilt.kharkov.ua},
O. E. Kvitnitskaya$^1$, I. K. Yanson$^1$, S.-L. Drechsler$^2$,
G. Behr$^2$, and S. Otani$^3$}

\affiliation{$^1$ B. Verkin Institute for Low Temperature Physics
and Engineering, National Academy  of Sciences of Ukraine,  47
Lenin Ave., 61103,  Kharkiv, Ukraine}

\affiliation{$^2$ Leibniz-Institut f\"ur Festk\"orper- und
Werkstofforschung Dresden e.V., Postfach 270116, D-01171 Dresden,
Germany}

\affiliation{$^3$ Advanced Materials Lab., National Institute for
Materials Sciences, 1-1 Namiki, Tsukuba, Ibaraki 305-0044, Japan}

\date{\today}

\begin{abstract}
The electron-phonon interaction (EPI) in transition metal
diborides $T$B$_{2}$ ($T$=Zr, Nb, Ta) is investigated by
point-contact (PC) spectroscopy. The PC EPI functions were
recovered and the EPI parameters $\lambda\lesssim 0.1$ were
estimated for all three compounds. Common and distinctive features
between the EPI functions for those diborides are discussed also
in connection with the superconductivity in MgB$_2$.

\end{abstract}

\maketitle

{\it Introduction}. The recent discovery of superconductivity in
MgB$_2$ at 39 K by  Akimitsu (see Nagamatsu {\it et al.}
\cite{Nagam})  renewed the interest in electron transport
measurements and activated a search for superconductivity in other
diborides. According to a recent review \cite{Buzea} no
superconducting transition has been observed so far in diborides
of transition metals $T$B$_2$ ($T$ = Ti, Zr, Hf, V, Cr, Mo). Only
NbB$_2$ is expected to superconduct with a rather low transition
temperature $T_c <$ 1 K and contradictory reports about
superconductivity up to $T_c$ = 9.5\,K in TaB$_2$ can be found in
the literature (see, e. g., Refs. 2, 3 and further Refs. therein).
Finally, the reported $T_c$ = 7 K in ZrB$_2$ \cite{Gaspa}
encourages further studies of these three diborides.

The goal of this paper is to determine the electron-phonon
interaction (EPI) function for selected diborides by means of
point-contact (PC) spectroscopy in order to address the above
mentioned issues about superconductivity in these compounds. The
measurement of the nonlinear conductivity of PC`s between two
metals allows us, in a direct way \cite{Yanson}, to recover the PC
EPI function $\alpha^2F(\omega)$. The knowledge of
$\alpha^2F(\omega)$ for conducting systems provides a consistent
check for the possibility of a phonon-mediated pairing mechanism,
e.g., by an estimation of the electron-phonon-coupling strength
characterized by the EPI parameter $\lambda = 2 \int \alpha^2
F(\omega)\omega^{-1}d\omega$. From a comparison of the
experimentally determined $\alpha^2F(\omega)$ with theoretical
calculations, different models and approaches can be
discriminated. Thus the PC spectroscopy could be helpful to
understand details of the EPI in the diborides under
consideration, and to evaluate contradictory reports about
possible superconductivity within this family.

{\it Experimental details}. We have used single crystals of
$T$B$_{2}$ ($T$=Zr, Nb, Ta) grown by the rf-heated floating-zone
method\cite{Otani}. Samples were prepared using a diamond wire
saw. The residual resistivity $\rho_0$ and the RRR of $T$B$_{2}$
are shown in Table I.

The experimental cell with the sample holders, which allows
mechanical movements of electrodes by differential screw
mechanism, was placed directly in liquid $^4$He to ensure good
thermal coupling. The PC`s were established {\it in situ} at low
temperatures by a touching of the cleaved surface of $T$B$_2$ by
the edge of another single crystals.
 We did not control the mutual orientation of electrodes;
therefore, the contact axis was not determined with respect to the
definite crystallographic direction. However, we did not observe
an appreciable variation of the maxima position (see additional
remarks for NbB$_2$ at the end of discussion below) and their
relative intensity in the spectra for different contacts.
Therefore, we believe the anisotropy is not crucial. Both the
differential resistance d$V/$d$I$ and the second derivative of the
$I-V$ characteristic d$^2V$/d$I^2(V)$ vs $V$ were registered using
a standard lock-in technique. The zero-bias resistance $R_0$ of
investigated contacts ranged from a few ohms up to several tens of
ohms at 4.2\,K.

\begin{table}
\begin{ruledtabular}
\caption{Parameters of investigated  $T$B$_2$ single
crystals.}\label{tab1}
\begin{tabular}{c|cccc}
 Samples\footnotemark[1] & $\rho_0\,\cdot10^9$, $\Omega$\,m & RRR &
 $n\footnotemark[2]\,\cdot 10^{-28}$, m$^{-3}$& \\
 \hline
 ZrB$_2$ & 3.31 & 24 & 13 & \\
 NbB$_2$ & - & - & 18.2 & \\
 TaB$_2$ & 220 & 1.2& 18.6 & \\
\end{tabular}
\end{ruledtabular}
 \footnotetext[1]{For NbB$_2$, the estimated $\rho$ was two
orders of magnitude larger compared to the other diborides
probably due to the presence of inner cracks. Therefore,
corresponding cells in the Table I are empty.}
 \footnotetext[2]{The density of carriers $n$ was estimated by the
number of valence electrons (4 for ZrB$_2$, 5 for NbB$_2$ and
TaB$_2$) per volume of the corresponding unit cell. }
\end{table}

{\it Results and discussion}. The voltage $V$ applied to the
ballistic contact defines the excess energy e$V$ of electrons;
therefore, for some of them backscattering processes caused by the
creation of phonons can take place. This results in a decrease of
the net current through the contact and leads to a nonlinear $I-V$
characteristic. According to the theory of Kulik, Omelyanchouk and
Shekhter\cite{KOS} in this case the second derivative
-\,d$^2I$/d$V^2(V)$ of the $I-V$ curve at low temperatures is
proportional to $\alpha_{\rm PC}^2\,F(\omega)$. In the free
electron approximation\cite{Kulik}
\begin{equation}
\label{pcs} -\frac{{\rm d}^2I}{{\rm d}V^2}\propto R^{\rm
-1}\frac{{\rm d}R}{{\rm d}V}= \frac{8\,{\rm e}d}{3\,\hbar v_{\rm
F}}\alpha_{\rm PC}^2(\epsilon)\,F(\epsilon)|_{\epsilon={\rm e}V} ,
\end{equation}
where $R={\rm d}V/{\rm d}I$,  $\alpha_{\rm PC}$, roughly speaking,
measures the interaction of an electron with one or another phonon
branch. The kinematic restriction of electron scattering processes
in a PC is taken into account by the factor
$K=1/2(1-\theta/\tan\theta)$, where $\theta$ is the angle between
initial and final momenta of scattered electrons [for transport
and Eliashberg EPI functions the corresponding factors are:
$K=(1-\cos\theta)$ and $K$=1, respectively]. Therefore in PC
spectra the large angle $\theta \to \pi$ scattering
(back-scattering) processes of electrons dominate.

From Eq.(\ref{pcs}) EPI function $\alpha_{\rm
PC}^2(\epsilon)\,F(\epsilon)$ can be expressed via the measured
rms signal of the first $V_1$ and second $V_2$ harmonics of a
small alternating voltage superimposed on the ramped dc voltage
$V$:
\begin{equation}
\label{pcs1} \alpha_{PC}^2(\epsilon)\,F(\epsilon)=
\frac{3\sqrt{2}}{4}\frac{\hbar v_{\rm F}}{{\rm
e}d}\frac{V_2}{V_1^2} .
\end{equation}
The PC diameter $d$, appearing in Eqs.(\ref{pcs}) and
(\ref{pcs1}), determines the constriction  resistance which
consists of  a sum of the ballistic Sharvin and the diffusive
Maxwell terms according to the simple formula
\begin{equation}
\label{Rwex} R_{\rm PC}(T) \simeq  \frac {16 \rho l}{3\pi d^2} +
\frac{\rho (T)}{d}
\end{equation}
derived by Wexler\cite{Wexler}, which is commonly used to estimate
the PC  diameter $d$. Here $\rho l = p_{\rm F}/n$e$^2$, where
$p_{\rm F}$ is the Fermi momentum, $n$ is the density of charge
carriers.

\begin{figure}
\begin{center}
\includegraphics[width=8cm,angle=0]{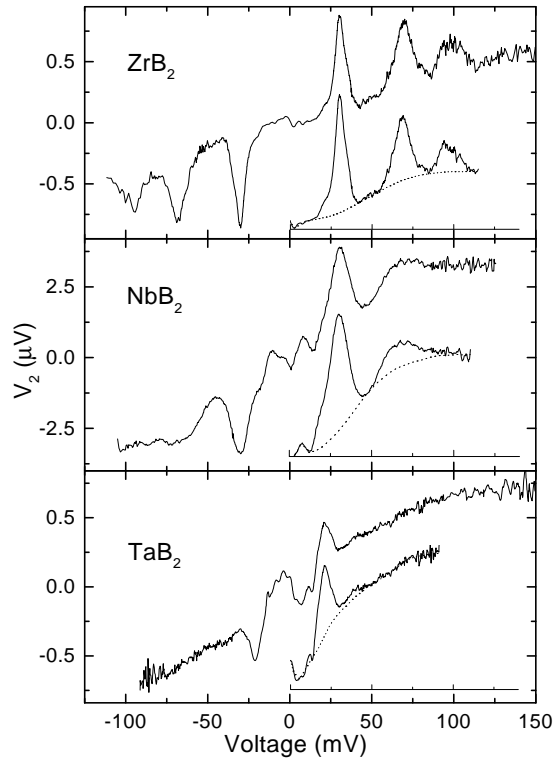}
\end{center}
\caption[] {Raw PC spectra d$^2V$/d$I^2(V)$ for investigated
compounds at $T=4.2$K.  The phonon structure is clearly resolved
with pronounced maxima up to 100\,mV (see ZrB$_2$), while for
TaB$_2$ only the low energy peak at 20\,mV is seen. The spectra
exhibit also a zero-bias anomaly, better pronounced in NbB$_2$ and
TaB$_2$.  The zero-bias resistance and modulation signal  for
ZrB$_2$ are $R_0=5.5\,\Omega$ and $V_1(0)$=0.8\,mV, for NbB$_2$
they are $R_0=50\,\Omega$, $V_1(0)$=2.8\,mV, and for TaB$_2$ they
are $R_0=25\,\Omega$, $V_1(0)$=1.3\,mV. The inset in each panel
shows the second derivative -\,d$^2I$/d$V^2\propto$
(d$^2V$/d$I^2$)(d$V$/d$I)^{-3}$ averaged for both polarities.
Dotted lines present thee behavior of the background. }
\label{fig1}
\end{figure}

Representative examples of measured d$^2V$/d$I^2(V)$ dependencies
are shown in Fig.\,1. Among tens of curves, which show
reproducible phonon structure for each compound, we selected
d$^2V$/d$I^2(V)$ characteristics with the most pronounced and
intensive maxima. A common feature for all crystals is the
presence of the main low energy maximum placed at about 30, 28 and
20\,mV for ZrB$_2$, NbB$_2$ and TaB$_2$, respectively. This is in
line with the common consideration that at fixed spring constants
the phonon frequency decreases with increasing atomic mass (see
Fig.\,2). Such a behavior suggests that the first peak corresponds
to the vibration of transition metal. Curiously, on the one hand,
the aforementioned peaks appear just above the maximal phonon
energy on PC spectra of the corresponding clean metals: Zr
\cite{Zr}, Nb and Ta from Ref.10 (see Table II). On the other
hand, the neutron data peak position for MgB$_2$ on Fig.\,2, at
about 36 meV \cite{Osborn}, is far below the straight line
connecting the $T$B$_2$ compounds. This might be considered a
consequence of a softening of the corresponding spring constants,
i. e. metallic bonds in MgB$_2$ instead of relatively strong
$T$$d$B$2p$ covalent bonds in the $T$B$_2$ series. Notice that the
recent data\cite{Bobrov} recovered maxima in the PC spectra of
MgB$_2$ even at lower energy of about 30 and 20\,meV. This cannot
be occasional. It points either to peculiarities in the lattice
dynamics of MgB$_2$ since such peaks are missing in standard
first-principle phonon calculations \cite{Kong,Heid} or to the
presence of other, nonphononic, low-frequency bosonic excitations
\cite{Shulga}. It is noteworthy that the presence of such
anomalous low-energy modes has also been noticed by other
experimental techniques: Raman scattering\cite{varelogiannis} and
tunneling measurements\cite{Dyachenko}. Their relationship to its
high critical temperature remains unclear at present.

\begin{figure}
\begin{center}
\includegraphics[width=8cm,angle=0]{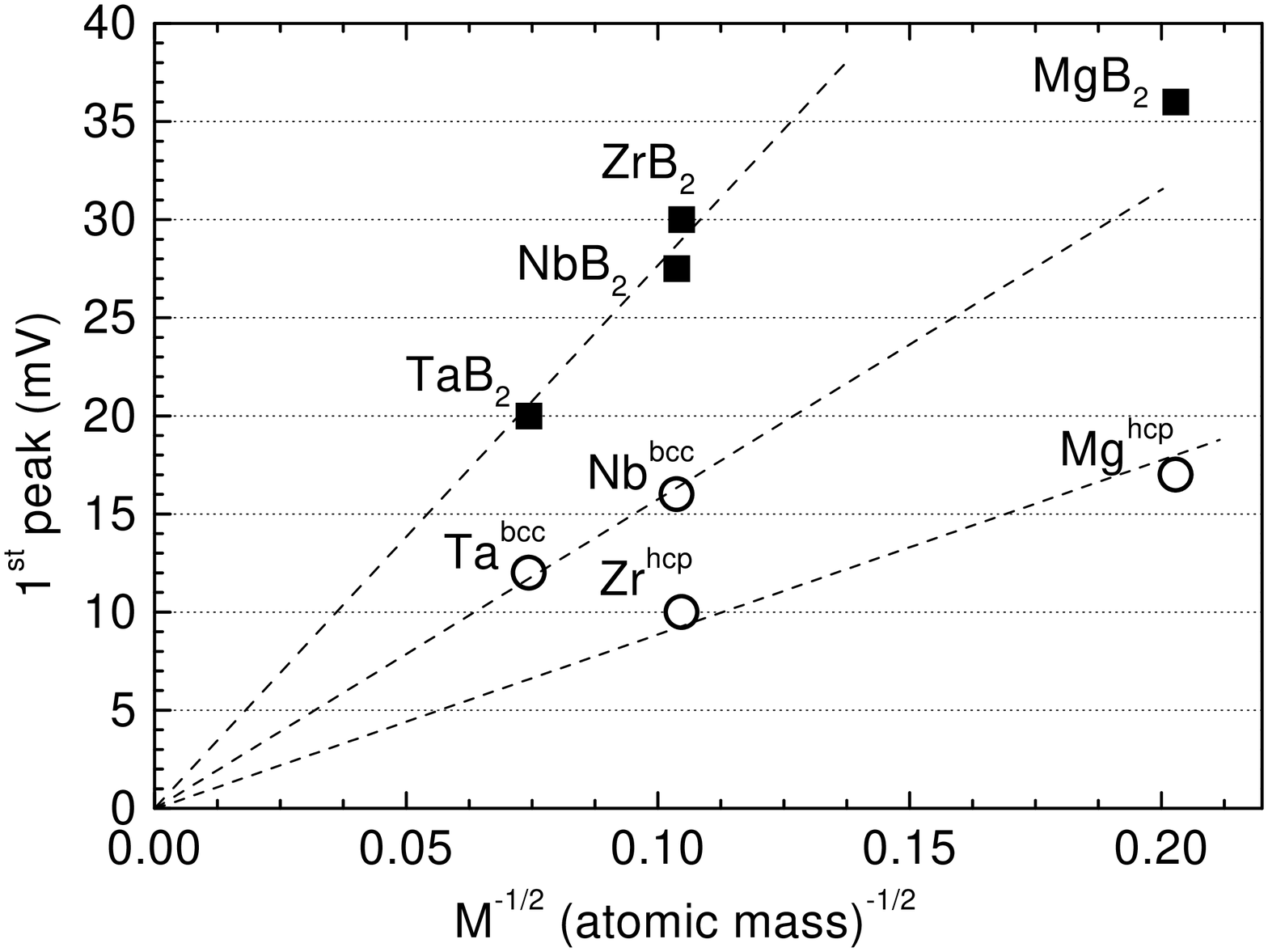}
\end{center}
\caption[] {The position of the first peak (squares) in the
measured PC spectra for ZrB$_2$, NbB$_2$ and TaB$_2$ vs the
inverse square root of atomic mass of the corresponding transition
metal. For MgB$_2$ the peak position is according to the inelastic
neutron scattering data\cite{Osborn}. Open circles show the
position of the first peak in the PC spectra for the corresponding
metals\cite{Zr,Khot}. Straight dashed lines are to guide eye. }
\label{fig2}
\end{figure}

Because of the large mass difference between the transition metal
and the boron atoms the boron derived modes are expected to
occurat much higher energy. Indeed, for ZrB$_2$ two additional
maxima at 70 and 100\,mV \cite{twophin} are well resolved, while
for NbB$_2$ the high energy part of the spectrum presents a broad
maximum around 60\,mV. For TaB$_2$ the high energy phonon peaks
were difficult to resolve, although according to a rough
estimation\cite{Rosner} the boron in-plane and out-of-plane
displacement modes should have energies of 98 and 85\,meV,
respectively. No spectral features were found for the
above-mentioned compounds above 100\,meV. This is in line with the
measured surface phonons\cite{Aizava} for ZrB$_2$ and NbB$_2$.
There  all phonon frequencies are below 100\,meV. Moreover, for
both compounds a phonon dispersion study\cite{Aizava} demonstrated
a gap between 30 and 50\,meV which separates acoustic and optic
branches. In this energy region a minimum in our PC spectra
occurs. Comparing the high energy parts of the ZrB$_2$ and the
NbB$_2$ PC spectra, we may support the statement\cite{Aizava}
that, for NbB$_2$ the boron surface phonon modes are softer and
more complex than in the case of ZrB$_2$. Possibly for this reason
in the PC spectra of NbB$_2$ all boron derived modes form broad
structureless maxima around 60-70 meV (see Fig.\,3).

\begin{table}
\begin{ruledtabular}
\caption{The phonon maxima and the EPI constant $\lambda$ in
$T$B$_2$ compounds measured by PC spectroscopy. The fifth column
shows the maximal energy for phonon features in the PC
spectrum\cite{Zr,Khot} for the corresponding transition metals:
$T$=Zr, Nb, Ta. }\label{tab2}
\begin{tabular}{c|ccccc}
Samples & 1$^{st}$ peak & 2$^{nd}$ peak & 3$^{d}$ peak
&$\hbar\omega_{max}^{{\rm T}}$& $\lambda_{\rm PC}$
\\
 & meV & meV &meV & meV& \\
 \hline
 ZrB$_2$ & $30\pm 0.5$ &  $68\pm 1$ &  $95\pm 2$ & 25& 0.06\\
 NbB$_2$ & $28\pm 2$ & $60\pm 5$ & - & 28& 0.08\\
 TaB$_2$ & $20\pm 1$ & 40{?} & - & 20 & 0.025\\
\end{tabular}
\end{ruledtabular}
\end{table}

To recover the spectral EPI function according to Eq.(\ref{pcs1}),
at first d$^2V$/d$I^2(V)$ has been transformed into
-\,d$^2I$/d$V^2(V)$, and thereafter the background (see Fig.\,1,
insets) was subtracted. There are a few models, both theoretical
and empirical, for the background behaviour. Using our experience
with PC spectra of different metals and compounds we have drawn
the background simply by eye with a dependence like
$\exp(-1/x^2)$, as shown in Fig.\,1, insets. During this procedure
we paid attention to the following: (i) zero-bias anomalies were
disregarded; (ii) the background curve is made to touch the
measured one at the minimum above the first peak, where a gap
between acoustic and optical phonons is expected; and (iii) above
100\,mV or below in the case of lacking visible maxima the
background coincides with the data. Fig.\,3 presents the recovered
PC EPI function for the investigated diborides calculated by
Eq.\,(2) with a Fermi velocity $v_{\rm F}= 1\times 10^{6}$m/s. To
obtain the PC diameter $d$ we used the Sharvin part of the PC
resistance from Eq.(\ref{Rwex}) and the calculated value $\rho l =
p_{\rm F}/n$e$^2=(3\pi^2)^{1/3}\hbar$e$^{-2}n^{-2/3}$ with $n$
from Table I. By virtue of the fact that $d \propto (\rho
l)^{1/2}\propto n^{-1/3}$ the simplicity of the evaluation of $n$
(see Table I footnote) is not crucial for the calculation of $d$
and afterward for estimation of the EPI parameter $\lambda$.

It is also seen from Fig.\,3 that the upper boundaries of the
ZrB$_{2}$ and NbB$_{2}$ spectra are at about 110\,meV and 90\,meV
(see Fig.\,3), which is much larger than their Debye temperatures
of 280 K and 460 K, respectively, estimated from the
Bloch-Gr\"uneisen temperature dependence of the
resistivity\cite{Gaspa}.

\begin{figure}
\begin{center}
\includegraphics[width=8cm,angle=0]{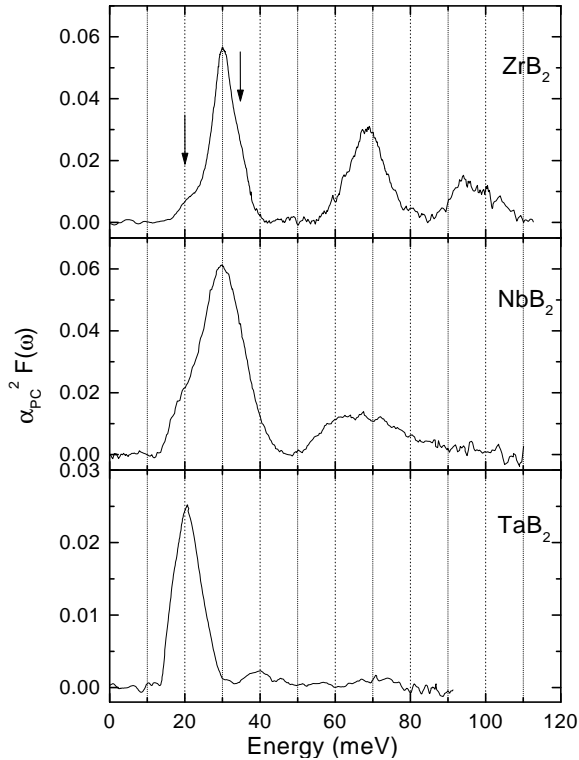}
\end{center}
\caption[] {The PC EPI function for ZrB$_2$, NbB$_2$ and TaB$_2$
recovered from the spectra in Fig.\,1. The vertical arrows mark
reproducible fine features as a bump and a shoulder for ZrB$_2$. }
\label{fig3}
\end{figure}

With the EPI function we have calculated the EPI parameter
$\lambda = 2 \int \alpha^2 F(\omega)\omega^{-1}d\omega$. We should
emphasize that, in general, due to presence of the $K$-factor,
$\lambda_{\rm PC} \neq \lambda_{Eliashberg}$; however, for many
superconductors it was found\cite{Khot} that $\lambda_{\rm PC}
\simeq \lambda_{Eliashberg}$. As we can see from Table II,
$\lambda_{\rm PC}$ is rather low for the diborides \cite{lambda}.
We remark that a small $\lambda_{\rm PC}\simeq$0.02 were also
reported for PC studies of the transition metal silicides NbSi$_2$
and TaSi$_2$ \cite{Balkas}. The reason for this can be the
deviation from the ballistic electron flow and the establishment
of a regime with $l_i \ll d$ in PC (here $l_i$ is the elastic mean
free path of electrons), which adds a prefactor\cite{KulYan} of
the order of $l_i/d$  in Eq.\,(\ref{pcs}). The large background
level and the absence of the high energy peaks, e. g., for the PC
spectra of TaB$_2$, may have the same origin: the lack of a
ballistic regime in our PC`s. Also note also, that the TaB$_2$
samples are of relative low quality (their RRR`s amounts to only
1.2) (see Table I); that is, a small $l_i$-value is already
expected in the bulk. Therefore, for TaB$_2$ we assume that
$\lambda$ is underestimated. However, in the case of ZrB$_2$,
where the PC spectra present distinct peaks up to the maximal
energy, and even at the first peak such details as a bump at
20\,mV and a shoulder at 35\,mV are seen, we believe that our
parameter $\lambda_{\rm PC}$ corresponds here to a real state of
the arts. Our results also show that NbB$_2$ has the largest
$\lambda_{\rm PC}$ among the studied compounds. Thus the search
for superconductivity in NbB$_2$ is more interesting. We should
note that, among the studied diborides, for NbB$_2$  we observed
the largest variation in the position of the first peak. For some
PC spectra the peak shifted down to 22-23\,mV. In this case any
high energy maxima, e. g., at 60\,mV, was difficult to resolve.
Most likely this is due to disturbed metal structure in the PC
area caused by low temperature deformation at the contact
formation. In any case the effect of anisotropy has to be studied.
Of course, to draw a more weighty conclusion about details of the
EPI and the $\lambda$ value in the presented $T$B$_2$ family, a
theoretical calculation of $\alpha_{\rm PC}^2 F(\omega)$ with the
mentioned $K$-factor, and a comparison with experimental data is
any desirable.

Finally, we should stress that by investigation of MgB$_2$ thin
films\cite{Bobrov} the EPI inelastic contribution to the PC
spectrum was estimated only in a few percent of the total PC
resistance, which is nearly an order of magnitude lower than for
the $T$B$_2$ under consideration. Why the EPI features on PC
spectra of MgB$_2$ are shallow and hardly reproducible is still
unclear at the moment. Possibly this is due to a weak EPI for a
3D-band of the Fermi surface sheet, which determines the PC
conductivity of thin films\cite{Yansmgb}, along with a small
$l_i/d$ ratio.

{\it Conclusion}. We have measured the PC spectra in transition
metal diborides: ZrB$_{2}$, NbB$_{2}$ and TaB$_{2}$. The spectra
exhibit structure up to an energy of about 100\,meV, which is
unequivocally caused by phonons. For all compounds the PC EPI
function was established and the EPI parameter $\lambda$ was
calculated. The obtained small $\lambda$ values strongly question
the reported bulk superconductivity in these compounds. The PC EPI
spectra of the above-mentioned diborides differs even
qualitatively from that measured earlier for superconducting
MgB$_2$.

{\it Acknowledgements}. The work in Ukraine was supported by the
National Academy of Sciences of Ukraine. The investigations were
carried out in part with the help of  equipment donated by the
Alexander von Humbold Stiftung (Germany). Further thanks to the
Deutsche Forschungsgemeinschaft for financial support. Discussions
with H. Rosner and S. V. Shulga are grateful acknowledged (S.-L.
D.).



\begin{thebibliography}{}

\bibitem{Nagam}
J. Nagamatsu {\it et al}.,
Nature (London) {\bf410}, 63 (2001).

\bibitem{Buzea} C.\,Buzea and T.\,Yamashita, Superconductors, Science \&
Technology, {\bf 14},  R115-R146 (2001).

\bibitem{Gaspa} V. A. Gasparov, N. S. Sidorov, I. I. Zver`kova, and
M. P. Kulakov, JETP Lett. {\bf 73}, 532 (2001).

\bibitem{Yanson} I. K. Yanson, Sov. J. Low Temp. Phys. {\bf 9}, 343 (1983).

\bibitem{Otani} S. Otani, M. M. Korsukova and T. Mitsuhashi, J. Crystal
Growth {\bf 186}, 582 (1998); {\bf 194}, 430 (1998).

\bibitem{KOS} I. O. Kulik, A. N. Omelyanchouk and R. I. Shekhter,  Sov. J.
Low Temp. Phys. {\bf 3} 840 (1977).

\bibitem{Kulik} I. O. Kulik,  Sov. J. Low Temp. Phys. {\bf 18}, 302 (1992).

\bibitem{Wexler} A. Wexler,  Proc. Phys. Soc. (London) {\bf 89}, 927
(1966).

\bibitem{Zr} PC spectra of Zr were measured by N. L. Bobrov and V. V. Fisun
(unpublished).

\bibitem{Khot}
A. V. Khotkevich and I. K. Yanson,  {\it Atlas of Point Contact
Spectra of Electron-Phonon Interaction in Metals} (Kluwer Academic
Publisher, Boston, 1995).

\bibitem{Bobrov} N. L. Bobrov, P. N. Chubov, Yu. G. Naidyuk {\it et al}.,
in {\it New Trends in Superconductivity}, edited by J. F. Annett
and S. Kruchinin, (Kluwer Acad. Publ., 2002) NATO Science Series
II: Mathematics, Physics and Chemistry,
Vol.67, p.225. 

\bibitem{Osborn}R. Osborn, E. A. Goremychkin, A. I. Kolesnikov, and
D. G. Hinks, Phys. Rev. Lett. {\bf 87}, 017005 (2001).

\bibitem{Kong}Y. Kong, O. V. Dolgov, O. Jepsen, and
O. K. Andersen, Phys. Rev. B {\bf 64}, 020501 (2001).

\bibitem{Heid}K. B.  Bohnen, R. Heid, and B. Renker, Phys. Rev.
Lett. {\bf 86}, 5771 (2001).

\bibitem{Shulga}S. V. Shulga {\it et al}.,
cond-mat/0103154.

\bibitem{varelogiannis}D. Lampakis {\it et al}.,
cond-mat/0105447.

\bibitem{Dyachenko}A. I. D'yachenko  {\it et al}.,
cond-mat/0201200.

\bibitem{twophin} It turns out that the 3$^{d}$ peak corresponds
to the sum of the 1$^{st}$  and the 2$^{nd}$ peaks (95\,mV
$\simeq$30\,mV + 68\,mV). However, our anylysis of two-phonon
contributions to the PC spectrum of ZrB$_2$ reject multiple phonon
nature of this maximum.

\bibitem{Rosner} H. Rosner {\it et al}.,
Phys. Rev. B {\bf 64}, 144516 (2001).

\bibitem{Aizava} T. Aizava, W. Hayami, and S. Otani, Phys. Rev. B
{\bf 65}, 024303 (2001).

\bibitem{lambda}$\lambda$ was calculated for a number of PC
spectra selected by commonly used in PC spectroscopy criteria
\cite{Yanson,Khot}. The scattering in $\lambda$ was within 20\%
more likely due to deviation from the ballistic regime. Table II
shows the maximal $\lambda$.

\bibitem{Balkas} O. P. Balkashin {\it et al}.,
Sol. State Commun. {\bf 100}, 293 (1996).

\bibitem{KulYan} I. O. Kulik and I. K. Yanson, Sov. J. Low Temp.
Phys. {\bf 4}, 596 (1978).

\bibitem{Yansmgb} I. K. Yanson {\it et al}., cond-mat/0206170.

\end{thebibliography}
\end{document}